\documentstyle[aps,prb]{revtex}

\begin{document}

\draft
\wideabs{
\title{Scattering on spin fluctuations in itinerant quantum 
disordered ferromagnets near quantum phase transition}
\author{Y. N. Skryabin}
\address{Institute for Metal Physics, 
Russian Academy of Sciences,
Ural Division, 
620219, Ekaterinburg, Russian Federation}
\author{A. V. Chukin}
\address{Ural State Technical University, 
620002, Ekaterinburg, Russian Federation}
\date{\today}
\maketitle

\begin{abstract}
Taking into account the long-ranged spin interactions due to 
weak localization effects in disordered itinerant ferromagnets 
the scattering of both electrons and neutrons on critical spin 
fluctuations near quantum phase transition is considered.
It is shown that for the case of low electron density, 
when $k_F \xi \ll 1$, where $k_F$ is the Fermi momentum
and $\xi$ is the magnetic correlation length, 
the transport cross section of electrons is proportional 
to the magnetic susceptibility as for the case of scattering 
by conventional critical fluctuations in absence of 
weak localization effects.
However, the intensity of the multiple small-angle neutron 
scattering becomes exponentionally small as compared with the 
intensity of the multiple scattering by conventional 
critical fluctuations.
\end{abstract}
\pacs{PACS numbers: 61.12.Bt, 64.60.Fr, 72.10.Di}
 
}
\section{Introduction}
It is well 
known \cite{Kirkpatrick96b,Vojta,Kirkpatrick97,Belitz97a,Belitz98}
in both disordered and clean itinerant quantum ferromagnets there 
are non-critical soft modes that exist in addition to the critical 
order parameter fluctuations. These soft modes couple to the 
critical modes and lead to an effective long-ranged 
interaction between the order parameter fluctuations. 
In disordered systems, they are the so-called ``diffusons''. 

According to \cite{Harris}, in the impurity system the correlation 
length exponent $\nu$, should satisfy the inequality $\nu \geq 2/d$, 
where $d$ is the dimension of the space. 
Thus, it is easy to see the mean-field value $\nu = 1/2$ 
does not satisfy this inequality for any dimension $d < 4$. 
In other words, the quenched disorder is a relevant perturbation 
with respect to mean-field theory. 

Recently Majumdar and Littlewood \cite{Majumdar} have studied 
transport in a low density electron gas coupled to ferromagnetic 
fluctuations near a finite temperature phase transition. 
They have suggested that this model describes ``colossal'' 
magnetoresistance in the pyrochlore 
Tl$_{2-x}$Sc$_x$Mn$_2$O$_7$, where the mechanism of ``colossal'' 
magnetoresistance is strongly different from the mechanism in the 
perovskite manganites La$_{1-x}$Sr$_x$MnO$_3$. 
The main assumption of this model is fact that magnetic 
moments are ordered ferromagnetically, independently from a low 
density electron gas.

Critical fluctuations usually lead to large scattering 
but the dominant long-ranged fluctuations near a ferromagnetic 
transitions have a negligible effect on transport because 
it is primarily modes near $2k_F$ which are effective in 
backscattering. 
The obvious and interesting exception is a low electron 
density system, $k_F a \ll 1$ ($k_F$ is the Fermi momentum 
and $a$ is the lattice constant), 
where the growth of magnetic fluctuations can be directly 
reflected in the resistivity \cite{Majumdar}.
 
The theory for the ``spin disorder'' contribution to 
resistivity near a ferromagnetic transition was given 
by de Gennes and Friedel \cite{deGennes-Friedel} 
and modified by Fisher and Langer \cite{Fisher-Langer}.

In this Letter we consider the peculiarities of scattering on 
spin fluctuations due to weak localization effects in disordered 
itinerant ferromagnets near the quantum phase transition at 
zero temperature.
It is shown that for the case of low electron density  
for $d=3$ the transport cross section of electrons 
is proportional to the magnetic susceptibility similarly
to the case of scattering by conventional critical 
fluctuations in absence of weak localization effects.
It occurs due to the fact that change of critical exponents 
upon the dimension of the space $d$ for the quantum phase 
transition conserves the critical exponent $\gamma$ 
for magnetic susceptibility.
However, the intensity of the multiple small-angle neutron 
scattering becomes proportional to $\exp (-L/l)$, 
where $L$ is the sample thickness and $l$ is the mean free path, 
as compared with the intensity of the multiple scattering 
by conventional critical fluctuations.

\section{Disordered quantum itinerant ferromagnet}

In disordered quantum ferromagnets the additional soft modes, 
``diffusons'', exist that cause the weak localization effects. 
These soft modes couple to critical modes and lead to an effective 
long-ranged interaction between the critical order parameter 
fluctuations \cite{Kirkpatrick96b}.
 
According to \cite{Kirkpatrick96b}, for small values of a wavenumber 
$|\bf q|$ and a bosonic Matsubara 
frequency $\Omega_n$ ($\Omega_n = 2\pi Tn$)   
the leading behavior of the order parameter correlation function 
$G({\bf q},\Omega_n)$ in disordered itinerant quantum ferromagnets is 
\begin{equation}
G({\bf q},\Omega_n) = \frac{\xi^{1/\nu}}{1 + 
a_{d-2}\xi^{1/\nu}|{\bf q}|^{d-2} + 
\xi^{1/\nu}{\bf q}^2 + a_{\omega}\xi^{1/\nu}\Omega_n/{\bf q}^2}.
\label{cf}
\end{equation}
Here $a_i$ are positive constants and $\xi \sim t^{-\nu}$ is the 
correlation length, where the $t$ is the distance from the 
critical point.
The most interesting contribution in Eq.\ (\ref{cf}) 
is the nonanalytic term $\sim |{\bf q}|^{d-2}$.
It is easy to see that due to this term there is the long-range 
interaction between the order parameter fluctuations  
which in real space is proportional to $r^{-2d+2}$. 
 
It should be noted here that this result occurs at zero temperature. 
At finite temperature an analytic expansion about $q=0$ exists  
and the conventional local functional for the free energy is 
obtained \cite{Kirkpatrick96b}.

It is well known from the theory of classical phase transitions 
that long-range interactions suppress fluctuations.
Using the renormalization group methods it has been 
found \cite{Kirkpatrick97} that only the Gaussian term 
is relevant in the free energy expansion for $d>2$.
In other words the effective long-range interaction between the order 
parameter fluctuations stabilizes the Gaussian fixed point, which 
describes the critical behavior of the system. 
 
For ${\bf q} = \Omega_n = 0$, the correlation function $G$ 
determines the magnetic susceptibility 
$\chi_m \sim G({\bf q} = 0,\Omega_n = 0)$ in zero field. 
Hence the critical exponent $\gamma$ has its usual mean-field 
value $\gamma = 1$. 
However, for non-zero wave-numbers the anomalous $|{\bf q}|^{d-2}$ 
term dominates the usual ${\bf q}^2$ dependence for all
dimensions $d < 4$. 
The correlation length exponent $\nu$ is given by
\begin{equation}
\nu = \cases{1/(d-2), &for $2<d<4$,\cr
              1/2, &for $d>4$.\cr} 
\label{nu}
\end{equation}
Note that $\nu \geq 2/d$, as it must be in general disordered 
systems \cite{Harris}. 
The wavenumber dependence of $G$ at $t=0$ is characterized 
by exponent $\eta$, which is defined as 
$G({\bf q},\Omega_n=0) \sim |{\bf q}|^{-2+\eta}$. 
From Eq. (\ref{cf}) we obtain 
\begin{equation}
\eta = \cases{4-d, &for $2<d<4$,\cr
               0,  &for $d>4$.\cr}
\label{eta}
\end{equation}

\section{Transport scattering cross section}

The transport relaxation rate $\tau ^{-1}$ is determined by 
\begin{equation}
\tau ^{-1} \propto \sigma _{\mbox{tr}},
\label{rate}
\end{equation}
where
\begin{equation}
\sigma _{\mbox{tr}} = 
\int _{0} ^{\pi} \sigma (q)\left( 1 - 
\cos \theta \right ) \sin \theta d \theta,
\label{transcross}
\end{equation}
is the transport scattering cross section.  
Here $\sigma (q)$ is the differential scattering cross section 
and $\theta $ the scattering angle.
The transfer momentum can be written as 
\begin{equation}
q = 2 k_{F} \sin (\theta /2) .
\label{momentum}
\end{equation}
Within the Ornstein-Zernike approximation we have 
\begin{equation}
\sigma (q) = g \xi ^{2} / ( 1 + q ^{2} \xi ^{2}),
\label{ornstein}
\end{equation}
where $\xi$ is the magnetic correlation length for finite 
temperature phase transition and $g$ is the square of 
the Born parameter of the magnetic scattering theory. 

The result for the transport scattering cross section is 
\begin{equation}
\sigma _{\mbox{tr}} = g \frac {1}{k_{F}^2} \left ( 4 - 
\frac {1}{k_{F}^2 \xi ^2} \ln ( 1 + 4 k_{F}^2 \xi ^2) \right ).
\label{ozrate}
\end{equation}
Note that for $k_{F} \xi \ll 1$ the transport cross section 
is proportional to $\xi ^2$. 

\section{Electron scattering in disordered itinerant ferromagnets}

Let us now discuss the scattering of electrons by critical order 
parameter fluctuations, which interact via dimensionally 
dependent long-range effective coupling.
According to Eq. (\ref{cf}), for $d > 4$ we have the case 
of electron scattering near finite temperature phase transition 
given by Eq.(\ref{ozrate}). 
However, for $2 < d < 4$ the differential scattering cross
section is   
\begin{equation}
\sigma ( q ) = g \xi ^{1/\nu} / ( 1 + a_{d-2} \xi ^{1/\nu} q^{d-2}),
\label{sigma}
\end{equation}
with the transport scattering cross section given by 
\begin{equation}
\sigma _{\mbox{tr}} = g \frac{8\xi^{1/\nu}}{d - 2} \int _{0} ^{1} 
\frac{z^{4\nu - 1}}{1 + bz} d z,
\label{imp}
\end{equation}
where $b = 2 k_{F} \xi a_{d-2}$.

Further, to be specific we consider the physical dimension 
$d = 3$. Eq. (\ref{imp}) can be written as
\begin{equation}
\sigma _{\mbox{tr}} = g \frac{8\xi}{b^4}\left [ \frac{b^3}{3} - 
\frac{b^2}{2} + b - \ln ( 1+b) \right ]. 
\label{3d}
\end{equation}
It is easy to see that for the case of the low electron density, 
when $k_{F} \xi \ll 1$, the transport relaxation rate for 3D 
is proportional to $\xi$ and the correlation length 
exponent $\nu = 1$. 
Nevertheless, this result is analogous to the Ornstein--Zernike 
approximation, where the transport cross section is proportional 
to $\xi^2$ and the correlation length exponent $\nu = 1/2$. 
So in the limit of small $q$ and $\Omega$ the transport 
cross section is proportional to $t^{-1}$ in general 
case due to the fact that the correlation function $G$ for this 
limit determines the magnetic susceptibility $\chi$ and its 
critical exponent $\gamma = 1$. 

Thus, we see that weak localization effects for the scattering 
of electrons by the critical spin fluctuations at the low 
electron density are irrelevant with respect to the scattering 
of electrons by conventional critical fluctuations.     

\section{Multiple small-angle neutron scattering on large-scale 
spin inhomogeneities}

It is well known that the small-angle neutron scattering is a 
widely used tool for studying of large-scale inhomogeneities in 
condensed matter. To determine the peculiarities of the critical 
order parameter fluctuations due to weak localization effects,  
we now discuss the multiple small-angle neutron scattering  
on spin fluctuations in disordered itinerant ferromagnet in 
vicinity of the quantum phase transition at zero temperature. 

Near the phase transition the scattering cross section on 
fluctuations increases and the mean free path $l$ decreases. 
In this case the multiple scattering of particles must be 
considered because the mean free path may be small as compared 
with sample thickness $L$. 
The theory of small-angle multiple scattering of particles 
(the so-called Moli\`{e}re theory) considered, for example, in  
Ref. \cite{Mott-Massey} has been generalized by Maleyev and Toperverg 
\cite{Maleyev80} on multiple scattering on critical fluctuations.
According to this theory the intensity of the scattering 
particles is given by the equation
\begin{equation}
I(q) = \frac{S}{2\pi} \int_0^\infty d \lambda \lambda 
J_0 \left ( \lambda \frac{q}{k} \right ) 
\exp \left [ -\frac{L}{l} \left ( 1 - \frac{\sigma_\lambda}
{\sigma_0}\right ) \right ],
\label{intensity}
\end{equation}
where $S$ is the sample area, $k$ is the momentum of incident 
particles and
\begin{equation}
\sigma_\lambda = \frac{2\pi}{k^2} \int_0^\infty d q q 
J_0 \left ( \lambda \frac{q}{k} \right ) \sigma (q).
\label{sigmal}
\end{equation}
Here $\sigma_0$ is the total cross section 
\begin{equation}
\sigma_0 = \frac{2\pi}{k^2} \int_0^{2k} \sigma (q) q d q.
\label{sigma0}
\end{equation}

Let us first consider scattering on conventional critical fluctuations. 
In the Ornstein-Zernike approximation the differential scattering 
cross section is given by Eq. (\ref{ornstein}). 
Then the total cross section is 
\begin{equation}
\sigma_0 = g \frac{2\pi}{k^2} \ln (2k\xi)
\label{total}
\end{equation}
and $\sigma_\lambda$ can be written in the form 
\begin{equation}
\sigma_\lambda = g \frac{2\pi}{k^2} \int_0^\infty 
J_0 \left ( \frac{\lambda}{k} q \right ) 
\frac{q}{\xi^{-2} + q^2} d q = g \frac{2\pi}{k^2} 
K_0 \left (\frac{\lambda}{k\xi} \right ),
\label{sl}
\end{equation}
where $K_0(z)$ is the modified Bessel function.
For small values of $\lambda /k \xi$ the leading behavior of 
$\sigma_\lambda$ is 
\begin{equation}
\sigma_\lambda =  - g \frac{2\pi}{k^2} \ln \left ( 
\frac{\lambda}{k\xi} \right ).
\label{sla}
\end{equation}
Using Eqs.(\ref{total}) and (\ref{sla}), the intensity of 
scattering can be written more explicitly, 
\begin{eqnarray}
I(q) & = & \frac{S}{2\pi} \int_0^\infty d \lambda \lambda 
J_0 \left ( \lambda \frac{q}{k} \right ) 
\exp \left [ -\frac{L \ln (2\lambda)}{l\ln (2k\xi)} 
\right ] \nonumber \\ 
& = & \frac{S}{2\pi} \int_0^\infty d \lambda \lambda 
J_0 \left ( \lambda \frac{q}{k} \right )(2\lambda)^{-s}, 
\label{inten}
\end{eqnarray} 
where $s = L/l\ln(2k\xi)$. 
From Eq. (\ref{inten}) we see that the range of small-angle 
scattering is
\begin{equation}
1 \ll \frac{L}{l} < 2\ln(2k\xi)
\label{range}
\end{equation}
and according to Ref. \cite{Maleyev80} 
\begin{equation}
I \sim \frac{S}{2\pi} \cases{(q/k)^{s - 2}, &for $q \gg 1/\xi$,\cr
                          (1/k\xi)^{s - 2}, &for $q \ll 1/\xi$.\cr} 
\label{mtres}			  
\end{equation}
At last, the mean free path $l = (n_0 \sigma_0)^{-1}$, where 
$n_0$ is the density of magnetic atoms, for this case can be 
written in the form 
\begin{equation}
l = \left ( n_0 g \ln (2k\xi) \right )^{-1}.
\label{path}
\end{equation}
Thus we see that for the small-angle scattering, when 
$k\xi \gg 1$, increasing of the correlation length $\xi$ 
near transition leads to decreasing of the mean free path. 

Now discuss the neutron scattering by the critical fluctuations 
in disordered systems taking into account weak localization 
effects. 
Again, to be specific we consider the dimension $d=3$, when 
at low $q$ the anomalous $|{\bf q}|^{d-2}$ term in 
Eq. (\ref{cf}) dominates the $q^2$ dependence.
Eq. (\ref{sigma0}) for $\sigma_0$ can be rewritten as,   
\begin{equation}
\sigma_0 = g \frac{2\pi}{k^2} \int_0^{2k} \frac{aq}{\xi^{-1} 
+ q} d q = g \frac{2\pi a}{k^2} \left [ 2k - \frac{1}{\xi}
\ln(1 + 2k\xi)\right ] 
\label{sigmaanom}     
\end{equation}
In general the small-angle scattering assumes that 
$k\xi \gg 1$. For this case we then have 
\begin{equation}
\sigma_0 \simeq g \frac{2\pi a}{k^2} \left [ 2k - 
\frac{1}{\xi}\ln(2k\xi) \right ].
\label{sigmaanomsa}
\end{equation}

Similarly, the equation for $\sigma_\lambda$ can be written as
\begin{eqnarray}
\sigma_\lambda & = & g \frac{2\pi}{k^2} \int_0^\infty J_0 
\left ( \frac{\lambda}{k}q \right ) \frac{aq}{\xi^{-1} 
+ q} d q \nonumber \\
& = & g \frac{2\pi a}{k^2} \left \{ \frac{k}{\lambda} - 
\frac{1}{\xi} \frac{\pi}{2} \left [ {\bf H}_0 \left (
\frac{\lambda}{k\xi} \right ) - 
Y_0 \left ( \frac{\lambda}{k\xi} \right ) \right ] 
\right \} \nonumber \\
& \simeq & g \frac{2\pi a}{k^2} \left \{ \frac{k}{\lambda} - 
\frac{1}{\xi} \left [ \frac{\lambda}{k\xi} - 
\ln \frac{\lambda}{k\xi} \right ] \right \},
\label{slambdaanom}
\end{eqnarray}
where ${\bf H}_0(z)$ is the Struve function, $Y_0(z)$ the Bessel 
function of the second kind and we use units where $a_1 = 1$. 

If we collect all contributions to the scattering intensity, 
we obtain
\begin{eqnarray}
I(q) & = & \frac{S}{2\pi} \int_0^\infty d \lambda \lambda 
J_0 \left ( \lambda \frac{q}{k} \right ) \times \nonumber \\ 
& & \exp \left [ - \frac{2 - (1/\lambda) - (1/k\xi) 
\ln (2\lambda) + \lambda/(k\xi)^2}{2 - (1/k\xi)\ln (2k\xi)} 
\frac{L}{l} \right].
\label{intenanom}
\end{eqnarray}
For $\lambda \gg k\xi$ the function $\sigma_\lambda$ 
decreases as $\lambda ^{-1/2}$ when $\lambda$ increases 
and the contribution to the intensity from $\lambda \gg k\xi$ is 
exponentially small for large values of $L/l$. 
Then the upper limit of integral over $\lambda$ in 
Eq. (\ref{intenanom}) must be smaller than $k\xi$.
Moreover, the main contribution to the integral gives the range 
$\lambda < k/q$, where the Bessel function 
$J_0 (\lambda q/k) \sim 1$, and for small-angle scattering 
the value of $\lambda$ must be bigger than 1. 
Now, it is easy to see that the intensity given by 
Eq. (\ref{intenanom}) is proportional $\exp (-L/l)$ both for 
$q \gg 1/\xi$ and $q \ll 1/\xi$.

\section{Conclusion}

We discussed the electron transport scattering cross section 
and neutron small-angle multiple scattering intensity 
on the critical order parameter fluctuations in disordered 
itinerant quantum ferromagnets with the low electron density. 
The crucial feature of quantum phase transitions 
at zero temperature is the coupling of the critical soft mode 
to the additional non-critical soft modes that leads 
to long-range interactions between critical order parameter 
fluctuations via these soft modes and thus considerably 
changes the critical behavior of the system.
In disordered systems, these soft modes are known as 
diffusons and cause weak localization effects. 
It is well known that long-range interactions suppress  
critical fluctuations.

Taking into account weak localization effects we obtained the 
transport cross section for low density electrons analogous 
to the Ornstein-Zernike approximation
 which describes the 
scattering of electrons by conventional critical fluctuations.
To understand this result one should remember that according 
to definition of the transport cross section the contribution 
of small angles is restricted by the factor $(1-\cos \theta)$. 
Hence, the dominant long-ranged fluctuations have a negligible 
effect on transport. 

In contrast, the intensity of the small-angle neutron scattering 
in disordered systems is determined namely by the long-ranged 
fluctuations.
So, in disordered systems, the scattering cross section 
is finite when the correlation length $\xi$ goes to infinity 
while for the conventional critical fluctuations one is diverged.
The intensity of the multiple small-angle scattering 
on 
large-ranged critical fluctuations is exponentially small. 
This is the result of suppressing of the critical fluctuations 
by the long-range interaction.
On the other hand, the scattering by the conventional critical 
fluctuations gives the contribution to the intensity of 
the multiple small-angle scattering given by Eq. (\ref{mtres}).
 
\acknowledgements

This work was partially supported by Russian Foundation for 
Basic Research (Project No. 97-02-17315).

\bibliography{data}
\bibliographystyle{prsty}
\end{document}